\documentclass[a4paper]{article}

\usepackage{INTERSPEECH2020}
\usepackage{graphicx}
\usepackage{cite}
\usepackage{multirow}

\title{The INESC-ID Multi-Modal System for the ADReSS 2020 Challenge}
\name{Anna Pompili$^1$, Thomas Rolland$^{1,2}$, Alberto Abad$^{1,2}$\thanks{This work has been partially supported by national funds through Funda\c c\~ao para a Ci\^{e}ncia e a Tecnologia (FCT) with reference UIDB/50021/2020 and by European Union funds through Horizon 2020 research and innovation programme under the Marie Sklodowska-Curie Grant Agreement No. 766287.}}
\address{
  $^1$INESC-ID, Lisbon, Portugal\\
  $^2$Instituto Superior T\'{e}cnico, Universidade de Lisboa, Portugal}
\email{anna.pompili@inesc-id.pt, thomas.rolland@hlt.inesc-id.pt, alberto.abad@inesc-id.pt}

\begin{document}

\maketitle
\begin{abstract}

This paper describes a multi-modal approach for the automatic detection of Alzheimer's disease proposed in the context of the INESC-ID Human Language Technology Laboratory participation in the ADReSS 2020 challenge. Our classification framework takes advantage of both acoustic and textual feature embeddings, which are extracted independently and later combined. Speech signals are encoded into acoustic features using DNN speaker embeddings extracted from pre-trained models. 
For textual input, contextual embedding vectors are first extracted using an English Bert model and then used either to directly compute sentence embeddings or to feed a bidirectional LSTM-RNNs with attention. 
Finally, an SVM classifier with linear kernel is used for the individual evaluation of the three systems. Our best system, based on the combination of linguistic and acoustic information, attained a classification accuracy of 81.25\%. Results have shown the importance of linguistic features in the classification of Alzheimer's Disease, which outperforms the acoustic ones in terms of accuracy. Early stage features fusion did not provide additional improvements, confirming that the discriminant ability conveyed by speech in this case is smooth out by linguistic data.
 
 \end{abstract}

\section{Introduction}
Alzheimer's Disease (AD), the most common cause of Dementia~\cite{WHOdementiaFS}, is a neurodegenerative disorder characterized by loss of neurons and synapses in the cerebral cortex. Its prevalence increases with age, a study on the U.S. census reported that 3\% of people aged 65-74, 17\% of people  aged 75-84, and 32\% of people  aged 85 and older have AD~\cite{hebert2013alzheimer}. As most countries are experiencing a general increase in average lifespan, it is expected a rapidly escalation of AD cases worldwide in the next thirty years~\cite{ADReport2015}. 
Pharmacological treatments may temporarily improve the symptoms of the disease, but they can not stop or reverse  its progression. For these reasons, there is an increasing need for additional, noninvasive, and cost-effective tools allowing a preliminary identification of AD in its early clinical stages.
Currently, AD is diagnosed through an analysis of patient clinical history and disability, neuropsychological tests, brain imaging and cerebrospinal fluid exams. 
Although the prominent symptoms of the disease are alterations of memory and of spatial-temporal orientation,  language impairments are also an important factor confirmed by current literature~\cite{reilly2011,kempler1995}.
Some of the most well known language impairments found in AD speech include naming~\cite{reilly2011}, word-finding difficulties~\cite{kempler1994}, repetitions~\cite{croisile1996}, an overuse of indefinite and vague terms~\cite{ahmed2013}, and inappropriate use of pronouns~\cite{ripich1988}.

Over the last years, there has been an increased interest from the research community in the automatic identification of AD through the analysis of speech and language abilities. 
Some studies have focused on syntactic or semantic features~\cite{hernandez2018,Fraser2016a}, some targeted plain acoustic approaches~\cite{konig2015automatic,haider2019assessment}, while other works have investigated a combination of temporal speech parameters and lexical measures~\cite{gosztolya2019,mirheidari2019}. Most of these approaches use handcrafted features and traditional classification algorithms. Very recent works investigated the use of automatically learned representations from deep neural networks
~\cite{chen2019attention,zargarbashi2019multi,warnita2018detecting,karlekar2018detecting}. Regardless of the approach used, the studies existing in the literature are difficult to analyze and compare due to the different datasets used.
In this scenario, the Alzheimer’s Dementia Recognition through Spontaneous Speech (ADReSS) challenge has been proposed, with the aim of providing researchers with a common, statistically balanced and acoustically enhanced dataset to test their approaches~\cite{luz2020alzheimer}.

In this work, we present  the multi-modal system proposed by the Human Language Technology Laboratory of INESC-ID for the ADReSS 2020 challenge. 
Our framework is designed to solve the task of automatically distinguishing AD patients from healthy individuals.
In our previous approaches to this topic~\cite{pompili2020pragmatic,pompili2018topic} we exploited lexical, syntactic, and semantic features with measures of local, global, and topic coherence, in order to provide a more comprehensive characterization of language abilities in AD and thus a more reliable identification.
In this work, we take the challenge of using automatically learned representations instead of traditional and consolidated handcrafted features, which already proven to achieve good classification results. Inspired by recent studies, we push the limit of deep neural models to work with extreme conditions, such the ones in the health domain, in which data scarcity is ordinary. Additionally, we combine both acoustic and linguistic information to have a complete picture of patient's disabilities, in a similar way to the type of information that clinicians receive during their interactions with patients. 

The rest of this work is organized as follows: Section \ref{section:SOA} introduces the relevant state on the art on the automatic identification of AD. Then, in Section \ref{section:corpus} and \ref{section:methods}, we present the dataset used in this study and a description of our methodology. Finally, classification results are reported in Section \ref{section:exp}, while conclusions are summarized in Section \ref{section:conclusions}.

\section{Related work}
\label{section:SOA}
The computational analysis of speech and language impairments in AD has gained growing attention in recent years.
Initially, existing studies explored engineered temporal and acoustic parameters of speech, linguistic features, or a combination of both.
K\"{o}nig \textit{et al.}~\cite{konig2015automatic} computed several temporal speech features on a dataset composed of 26 AD and 15 healthy subjects, while performing different tasks of isolated and continuous speech. By considering different features according to the task, the authors achieved an accuracy of 87\% in the automatic identification of AD. 
Fraser \textit{et al.}~\cite{Fraser2016a} used more than 350 features to capture lexical, syntactic, grammatical, and semantic phenomena from the transcriptions of a picture description task. With a selection of 35  features, the authors achieved a classification accuracy of 81.92\% in distinguishing individuals with AD from healthy controls. Pompili \textit{et al.}~\cite{pompili2020pragmatic} exploited lexical, syntactic, semantic and pragmatic features from the descriptions of the Cookie Theft picture~\cite{Goodglass2001} attaining an accuracy of 85.5\% in the task of classifying AD patients.
Gosztolya \textit{et al.}~\cite{gosztolya2019} collected a dataset composed of 75 Hungarian speakers (25 AD, 25 MCI, and 25 healthy subjects) performing two tasks eliciting continuous speech. The set of features used considered demographic attributes, acoustic and linguistic features. Using only acoustic or linguistic information the authors achieved an accuracy of 82\% in distinguishing AD patients from healthy subjects. When the two types of features were combined, the accuracy increases to 86\%.

More recently, researchers are shifting their focus towards more complex architectures capable of overcoming the limitations of traditional approaches.
Warnita \textit{et al.}~\cite{warnita2018detecting} proposed an approach relying only on acoustic data computed from continuous speech and gated Convolutional Neural Network (GCNN). Using majority voting on speaker and the Paralinguistic Challenge (IS2010) feature set, the authors achieved an accuracy of 73.6\%.  
Karlekar \textit{et al.}~\cite{karlekar2018detecting}, on the other hand, investigated linguistic impairments using CNN, LSTM-RNNs, and a combination of both. In this way, they obtained an accuracy of 91.1\% in classifying AD patients. 
Chen \textit{et al.}~\cite{chen2019attention} went further, proposing a network based on attention mechanism and composed of a CNN and GRU module. In this way, the architecture should be able to analyze both local speech patterns and global macro-linguistic functions. The accuracy achieved in distinguishing AD patients was of 97.42\%.
Finally, Zargarbashi \textit{et al.}~\cite{zargarbashi2019multi} designed a multi-modal feature embedding approach based on $N$-gram, \textit{i-vectors}, and \textit{x-vectors}. 
Classification accuracy results achieved with each of these models were, respectively, of 78.2\%, 75.9\%, and 75.1\%. The joint fusion of the three models reached an accuracy of 83.6\%.

Our work differs from previous studies for several reasons. First, to process the text data, we use contextual embeddings vectors as input to two different systems. One based on the training of a Global Maximum pooling and a bidirectional LSTM-RNNs architectures, and one based on the statistical computation of sentence embeddings. The latter presents the advantage of being a simple approach,  which does not require the training of deep, data-demanding architectures. Second, for the audio recordings, we use DNN speaker embeddings extracted from pre-trained models. These learned, speaker representative vectors have recently shown their potential in the discrimination of neurodegenerative disorders
~\cite{botelho2020pathological}. To the best of our knowledge, this is the first work that jointly uses automatically learned representations from neural models, instead of engineered features, for both audio signals and textual data. In fact, although existing studies have shown that linguistic impairments in AD appear to be more important than acoustic ones, traditional literature provide convincing evidence that using both source of information will definitively improve the accuracy of automatic diagnosis methods.

\section{Corpus}
\label{section:corpus}
The ADReSS dataset contains the speech recordings and corresponding annotated transcriptions of 156 subjects, 78 AD patients, and 78 healthy control matched for age and gender. Data were divided into two partitions, training and test sets composed of 108 and 48 subjects, respectively. Recorded participants were required to provide the descriptions of the Cookie Theft picture from the Boston Diagnostic Aphasia Examination~\cite{Goodglass2001}. Speech recordings were segmented using Voice Activity Detection (VAD) and later normalised~\cite{luz2020alzheimer}. The dataset  contained both full enhanced audio, and normalised audio chunks.

In our approach, we have used both the full enhanced audio and the transcriptions. 
The latter were annotated with disfluencies, filled pauses, repetitions, and other more complex events. 
However, to build an automated system requiring a minimal annotation effort, we removed all the annotations not corresponding to the plain textual representation of words, thus, better resembling the output that can be generated by an Automatic Speech Recognition (ASR) system.
Overall, the whole set of transcriptions contained 17127 words, of which 1009 were unique. More detailed information about the duration and size of the ADReSS dataset are reported in Table~\ref{tab:corpus}.

\begin{table}[t]
\caption{Statistical information on the ADReSS dataset}
\label{tab:corpus}
\resizebox{\columnwidth}{!}{
\begin{tabular}{lccc}
\hline
                        & \multicolumn{2}{c}{\textbf{Train}}                         & \multicolumn{1}{c}{\textbf{Test}} \\ \hline
                        & \multicolumn{1}{c}{Control} & \multicolumn{1}{c}{AD} & \multicolumn{1}{c}{--}            \\
\textbf{Audio Full}     & 00:55:46                    & 01:14:00                     & 01:06:00                          \\
\textbf{Audio chunks}   & 00:30:11                    & 00:26:31                     & 00:26:32                             \\
\textbf{\# words (unique)}     & 6097 (567)            & 5494 (552)           & 5536 (602)                  \\ \hline   

\end{tabular}
}
\end{table}

\section{Proposed methods}

\label{section:methods}
\begin{figure}
\includegraphics[width=0.5\textwidth]{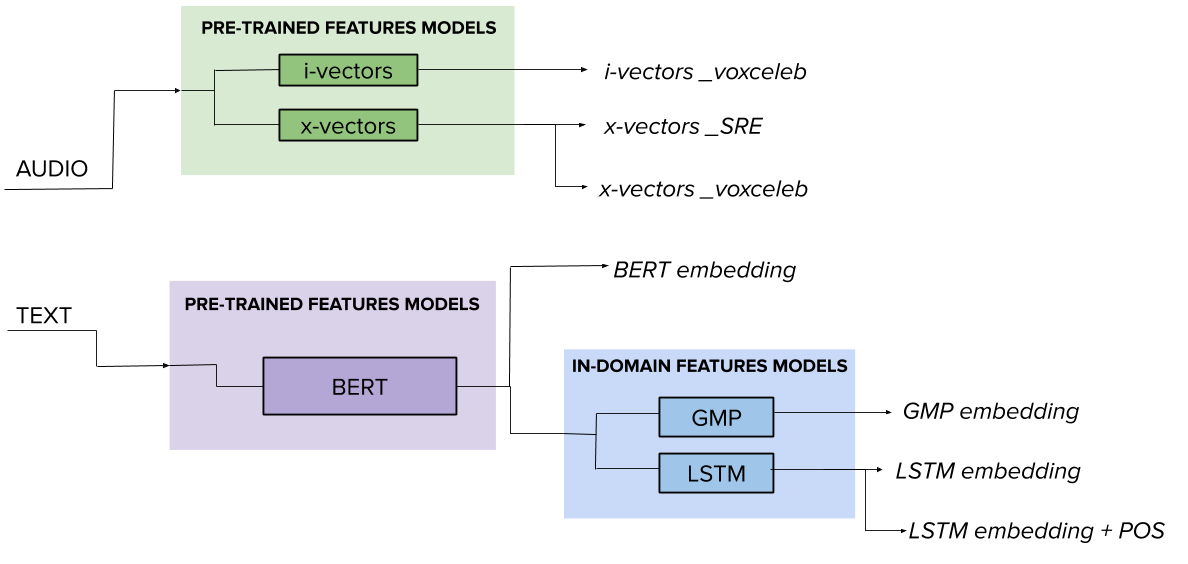}
\caption{Summary of embedding-based approaches}
\label{fig:blocks}
\end{figure}

As shown in Figure \ref{fig:blocks}, our multi-modal framework is based on the independent generation of acoustic and textual feature embeddings. Then, we perform an early fusion of the output of the two systems to obtain a single feature vector  containing a compact representation of both speech and language characteristics. Finally, classification is performed with an SVM classifier with linear kernel. The two systems are described in the remainder of this section. 

\subsection{Acoustic system}
The acoustic system is strongly based on two models borrowed from the speaker verification field, \textit{i-vectors}~\cite{dehak2010front} and \textit{x-vectors}~\cite{snyder2017deep}. \textit{i-vectors} are statistical speaker representation vectors that have been recently used for the classification of Parkinson's Disease and for the automatic prediction of dysarthric speech metrics~\cite{hauptman2019identifying,laaridh2017automatic}. \textit{X-vectors} are discriminative deep neural network-based speaker embeddings that have outperformed \textit{i-vectors} in speaker and language recognition tasks~\cite{snyder2018x,snyder2018spoken,snyder2017deep} and have been successfully applied to AD, obstructive sleep apnea and pathological speech detection~\cite{perero2019modeling,botelho2020pathological}. Both models allow to extract a fixed sized feature vector from variable length audio signal.

Taking into consideration the small size of the ADReSS dataset, we preferred to exploit already existing pre-trained models to produce our acoustic feature embeddings, rather than  training them using in-domain challenge data.
To this end, for the \textit{x-vectors} framework we use both the SRE and the Voxceleb models. 
The first was trained mainly on telephone and microphone speech using data from the Switchboard corpus, Mixer 6, and NIST SREs~\cite{snyder2018x}. 
The latter was trained on augmented VoxCeleb 1 and VoxCeleb 2 
datasets,
which contains speech from speakers spanning a wide range of different ethnicities, accents, professions and ages.~\cite{snyder2018x,nagrani2017voxceleb}. This dataset was used also to build the \textit{i-vectors} pre-trained model used in this work. 

The inputs to the pre-trained SRE and Voxceleb models consisted of 23 and 30-dimensional MFCC vectors, extracted with Kaldi~\cite{povey2011kaldi} from the full recordings, using default values for window size and shift. Non-speech frames were removed using energy-based VAD.
For the \textit{x-vectors} model, the last layers of the pre-trained model, before the softmax output layer, can be used to compute the embeddings. In this work, we extracted a 512-dimensional \textit{x-vectors} at layer \textit{segment6} of the network.

The \textit{i-vectors} models, is based on GMM-UBM. The universal background model (UBM) is used to capture statistics about intra-domain and inter-domain variabilities and a projection matrix is used to compute \textit{i-vectors}. We extracted a 400-dimensional \textit{i-vectors}. 

\subsection{Linguistic system}
We followed two different approaches to obtain textual feature embeddings. First, we investigated the feasibility of training deep architectures with a corpus of reduced dimension like the one used in this challenge. Then, this method is compared with a less data-demanding one, based on the statistical computation of sentence embeddings using a pre-trained model. 
Both strategies rely on contextual word embeddings as input, but they provide different types of learned representations as output.
In fact, to combine the information from the linguistic and the acoustic systems, the trained architectures are used only to extract linguistic features from the last layer of the models, before the final classification. In this way, we obtain a single 768-dimensional feature vector for an entire description. The sentence embedding approach, on the other hand, provide a single 768-dimensional vector for each sentence of a description. These features are then  used to classify between AD patients and healthy subjects.
For both approaches, the first step of the pipeline deals with the normalization of the data provided in the ADReSS dataset. In fact, we recall that besides the plain transcription of the descriptions these also contain additional annotations and information that were removed.
Then, we encode each word of the clean transcriptions into a 768-dimensional context embedding vector using a frozen English Bert model pre-trained with 12-layers, 768-hidden. This representation is fed to our two linguistic systems, described hereafter.

The first system is derived from the ComParE2020 Elderly Challenge baseline~\cite{schuller2020interspeech}, and was obtained by adapting the original code to deal with the classification of AD.
With this approach, three different neural models are trained on top of contextual word embeddings: (i) a Global Maximum pooling, (ii) a bidirectional LSTM-RNNs provided with an attention module, and (iii) the second model augmented with part-of-speech (POS) embeddings.
During training, the loss is evaluated on the development set.

The second system provides the advantage of not requiring an additional phase of model training. Similarly to the approach followed with the acoustic system, we use automatically learned representations extracted from a pre-trained model to directly characterize linguistic deficits in AD. The contextual word embeddings obtained for each word of the clean transcriptions are now used to compute an embedding vector of fixed size for each sentence of a description. Sentence embeddings were successfully employed in tasks of humor detection and more generally sentiments analysis \cite{annamoradnejad2020colbert, le2014distributed}  and information retrieval \cite{le2014distributed}. In our approach, sentence embeddings are computed by averaging the second to twelfth hidden layers of each word. 

\begin{table}[t]
\caption{Results of different acoustic approaches on the development set}
\label{tab:res_dev_ac}
\resizebox{\columnwidth}{!}{
\begin{tabular}{lcccc}
\hline
                                                & \multicolumn{1}{l}{\textbf{Accuracy}} & \multicolumn{1}{l}{\textbf{Precision}} & \multicolumn{1}{l}{\textbf{Recall}} & \multicolumn{1}{l}{\textbf{F1 Score}} \\
                                                \hline
\textbf{\textit{\textbf{x-vectors}\_Vox}   }        & 0.6818                               & 0.6834                                & 0.6919                             & 0.6812                               \\
\textbf{\textit{\textbf{x-vectors}\_SRE}}                & \textbf{0.7273}                               & \textbf{0.7273}                                & \textbf{0.7273}                             & \textbf{0.7273}                               \\
\textbf{\textit{\textbf{i-vectors}\_Vox} }          & 0.6818                               & 0.7292                                & 0.6818                             & 0.6645                               \\
\textbf{\textit{i-vectors}\_Vox\_\textit{x-vectors}\_Vox} & 0.7273                               & 0.7273                                & 0.7273                             & 0.7273                               \\
\textbf{\textit{i-vectors}\_Vox\_\textit{x-vectors}\_SRE}      & 0.7273                               & 0.7351                                & 0.7273                             & 0.7250                                \\
\hline
\end{tabular}
}
\end{table}

\section{Results and discussion}
\label{section:exp}
The ADReSS dataset contains only training and test partitions and for the latter the ground truth is not provided. Thus, in order to test our approaches, we retain the 20\% of the data from the training set and use it as development set. In this way, we are left with 86 subjects for training, 22 for development, and 48 for testing. While creating the additional partition, we kept the dataset gender balanced.

As briefly mentioned, our evaluation method relies on SVM~\cite{cortes1995svm} with linear kernel, based on a liblinear implementation. The complexity parameter \textit{C} was optimised during the development phase. 
The results reported in Tables~\ref{tab:res_dev_ac} and \ref{tab:res_dev_ling} are obtained using the best complexity configuration.
Features were normalized to have zero mean and unit variance.
In the remainder of this section we first describe our results on the development set for each system independently and then for their final fusion. Finally, for the best systems, we report the results obtained on the test set.

\subsection{Results on the development set}
\subsubsection{Acoustic system}
Results using different automatically learned acoustic features embeddings are summarized in Table~\ref{tab:res_dev_ac}. Also in this case, we explored different independent models and then we do an early fusion of the best acoustic results attained.
From Table~\ref{tab:res_dev_ac} is possible to observe that the \textit{x-vectors} Voxceleb model usually achieve a lower classification accuracy. However, when we combine both \textit{i-vector}s and \textit{x-vectors} extracted from this model, the accuracy resulting from their fusion is comparable to that of \textit{x-vectors} using the SRE model, which is currently our best result on the development set.
These outcomes are slightly lower than those found in the literature review for similar works. 
In fact, we recall that Warnita \textit{et al.}~\cite{warnita2018detecting} and Zargarbashi \textit{et al.}~\cite{zargarbashi2019multi} obtained an accuracy of 73.6\%,  75.9\%, and 75.1\%, using, respectively a gated CNN with the IS10 acoustic feature set and the \textit{i-vectors}/\textit{x-vectors} paradigms.
Our approach, however, is different from the ones of these authors since we are using a smaller dataset and do not rely on DNN training. Nevertheless, since we are interested in corroborating these results on the test set, we select the acoustic feature embeddings extracted from the pre-trained \textit{x-vectors} SRE model for the evaluation.

The use of pre-trained acoustic embedding extractors has been motivated by the reduced size of the ADReSS dataset, that we considered to be insufficient for data hungry deep learning approaches. To confirm this, we also trained an end-to-end LSTM model for AD classification. The architecture consisted of one dense and two LSTM layers with a softmax activation function. The network took as input chunks of 500 voiced frames using 23-dimensional MFCC with delta and delta-delta. Majority voting was performed over all the chunks from the same speaker to generate a single prediction per speaker.
This end-to-end approach performed very poorly, with an accuracy around chance result in the development set, confirming our expectations that the ADReSS dataset is not suited for training a deep learning end-to-end system.

\begin{table}[t]
\caption{Results of different linguistic approaches on the development set}

\label{tab:res_dev_ling}
\resizebox{\columnwidth}{!}{
\begin{tabular}{lcccc}
\hline
                                   & \multicolumn{1}{l}{\textbf{Accuracy}} & \multicolumn{1}{l}{\textbf{Precision}} & \multicolumn{1}{l}{\textbf{Recall}} & \multicolumn{1}{l}{\textbf{F1 Score}} \\ \hline
\textit{\textbf{Global Max Pool.}} & 0.7727                               & 0.7947                                & 0.7728                             & 0.7684                               \\
\textit{\textbf{LSTM-RNNs}}        & 0.8182                               & 0.8182                                & 0.8182                             & 0.8182                               \\
\textit{\textbf{LSTM-RNNs Pos}}    & 0.8636                               & 0.8667                                & 0.8637                             & 0.8634                               \\
\textit{\textbf{GMax/LSTM-RNNs/LSTM-RNNs-Pos} }                   & \textbf{0.9091}                               & \textbf{0.9091}                                & \textbf{0.9091}                             & \textbf{0.9091}  \\  
\textit{\textbf{Sentence emb. - maj. vote}}                    & 0.7727                               &0.7947                & 0.7728                           & 0.7684
\\ \hline
\end{tabular}
}
\end{table}

\subsubsection{Linguistic system}
Results obtained with our different linguistic systems are summarized in Table~\ref{tab:res_dev_ling}. This table reports the performance for the features trained with the three neural models, their fusion, and finally for the sentence embeddings approach. For the latter, we present only results achieved using a majority voting over the entire description.
Our best classification result attained an accuracy of 90.91\% on the development set using the fusion of the linguistic features sets generated by the three neural models. Comparing this result with the one obtained by sentence embeddings, we acknowledge that neural models outperform simpler strategies even with constrained training data. This was somehow surprising and in contradiction with similar experiments performed with the acoustic system. We hypothesize that the large amount of contextual information provided by the Bert model is helpful in overcoming the limited size of the ADReSS dataset. Nevertheless, we suspect that the high accuracy attained with neural models may be too optimistic, due to the fact of having used the development set both for testing and evaluating the model's loss. Thus, in spite of their lower outcome, the sentence embeddings approach is selected as one of the systems to be evaluated on the test set. In fact, on the one hand, we think that they may represent a more reliable system, since do not require additional training. On the other hand, we also observe that they achieve higher classification scores, when compared with a similar approach based on GloVe embeddings
~\cite{mirheidari2018detecting}, thus corroborating our decision.

\subsubsection{Fusion of systems}
To provide a comprehensive evaluation of speech and language impairments in AD, the best results obtained with both the acoustic and the linguistic systems where combined together in an early fusion fashion.
We merged the \textit{x-vectors} features set obtained with the SRE model with the combination of linguistic feature sets (GMax/LSTM-RNNs/LSTM-RNNs-Pos) generated by the three neural models. Unfortunately, results on the development set using this extended set of features did not provide any further improvements with respect to using the linguistic system alone. We believe that, in this case, the predictive ability of linguistic features completely override acoustic ones.
Nevertheless, we select the combination of these two systems as our main system for the evaluation.

\subsection{Results on the test set}
Overall, we submitted three systems for the evaluation: (i) the fusion of the best results achieved by the linguistic and acoustic systems, (ii) sentence embeddings, (iii) the best acoustic system. A summary of these results is reported in Table~\ref{tab:res_test}. 
In general, we found a consistent impoverishment of the performance of our methods when evaluated on the test set, 
even for those systems based on features that do not required a training phase.
The first system submitted achieved the best result, with an accuracy of 81.25\%, showing that the use of deep architectures with contextual word embeddings are actually able of overcoming the limitation of a constrained dataset. The worse result is achieved by the acoustic system alone, with an average accuracy of 54.17\%. This outcome is lower than the one found in the ADReSS baseline (62.50\%)~\cite{luz2020alzheimer}, indicating that there is still room for improving our acoustic approach.
We relied on pre-trained models to overcome the lack of data, but we ended up with a similar problem. It is likely the case that an adaptation of these models to the characteristics of elderly speech would allow for better performance.

\begin{table}[t]
\caption{Results of different acoustic and linguistic approaches on the test set}
\label{tab:res_test}
\resizebox{\columnwidth}{!}{
\begin{tabular}{llcccc}
\hline
                                     & \textbf{Class} & \multicolumn{1}{l}{\textbf{Accuracy}} & \multicolumn{1}{l}{\textbf{Precision}} & \multicolumn{1}{l}{\textbf{Recall}} & \multicolumn{1}{l}{\textbf{F1 Score}} \\ 
                                     \hline
\textit{\textbf{Fusion of system}}   & AD             & \multirow{2}{*}{\textbf{0.8125}}                     & 0.9412                                 & 0.6667                              & 0.7805                                \\
                                     & non-AD         &                                       & 0.7419                                 & 0.9583                              & 0.8364                                \\
\textit{\textbf{Sentence embedding}} & AD             &  \multirow{2}{*}{\textbf{0.7292}}                                    & 0.8235                                 & 0.5833                              & 0.6829                                \\
                                     & non-AD         & \multicolumn{1}{l}{}                  & 0.6774                                 & 0.8750                              & 0.7636                                \\
\textit{\textbf{x-vectors\_SRE}}     & AD             & \multirow{2}{*}{\textbf{0.5417}}                                      & 0.5417                                 & 0.5417                              & 0.5417                                \\
                                     & non-AD         & \multicolumn{1}{l}{}                  & 0.5417                                 & 0.5417                              & 0.5417                                \\ \hline
\end{tabular}
}
\end{table}

\section{Conclusions}
\label{section:conclusions}
In this work we presented a multi-modal approach to the classification of AD based on automatically learned feature representations. Both for the acoustic and linguistic systems, we investigated feature embedding vectors extracted from pre-trained models, as well as the feasibility of training deep neural architectures. Using a combination of both approaches, we attained an accuracy of 90.91\% and 81.25\% on the development and test sets, respectively. Our results showed that acoustic systems, in comparison to linguistic ones, require more data in order to improve the predictive ability of neural models and  obtain fine-tuned features representations. 
Nonetheless, it is worth noting that  linguistic systems used manually generated transcriptions.
In the presence of potential ASR errors --which are commonly exacerbated in the case of atypical speech, such as AD speech--,  acoustic systems may play a more relevant role. The impact of these errors could be an interesting analysis for future work, as well as the investigation of robust acoustic methods and models specially tailored to the elderly and AD speech characteristics.

\bibliographystyle{IEEEtran}

\bibliography{mybib}

\end{document}